\documentclass[sigconf]{acmart}
\usepackage{balance}
\AtBeginDocument{%
  }

\copyrightyear{2026}
\acmYear{2026}
\setcopyright{cc}
\setcctype{by}
\acmConference[UbiComp Companion '26]{Companion of the 2026 ACM International Joint Conference on Pervasive and Ubiquitous Computing}{October 11--15, 2026}{Shanghai, China}
\acmBooktitle{Companion of the 2026 ACM International Joint Conference on Pervasive and Ubiquitous Computing (UbiComp Companion '26), October 11--15, 2026, Shanghai, China}
\acmDOI{10.1145/3798063.3837138}
\acmISBN{979-8-4007-2533-3/2026/10}

\begin{document}
\raggedbottom
\title{“Black Mirror?”: Public Sensemaking of AI-Powered Lifelogging Wearables}

\author{Ying Ma}
\orcid{0000-0001-5413-0132}
\email{ying.ma1@student.unimelb.edu.au}
\affiliation{%
   \department{School of Computing and Information Systems}
  \institution{University of Melbourne}
  \city{Melbourne}
  \country{Australia}}

\author{Jarod Govers}
\email{jarod.govers@unimelb.edu.au}
\orcid{0000-0002-7648-318X}
\affiliation{%
  \department{School of Computing and Information Systems}
  \institution{University of Melbourne}
  \city{Melbourne}
  \country{Australia}}

\author{Le Fang}
\email{le.fang2@unimelb.edu.au}
\affiliation{%
    \department{School of Computing and Information Systems}
  \institution{University of Melbourne}
  \city{Melbourne}
  \country{Australia}}

\author{Shuning Zhang}
\orcid{0000-0002-4145-117X}
\email{zsn23@mails.tsinghua.edu.cn}
\affiliation{%
    \department{Institute for Network Sciences and Cyberspace}
  \institution{Tsinghua University}
  \city{Beijing}
  \country{China}
}

\author{Yongquan ‘Owen’ Hu}
\email{yongquanhu.work@gmail.com}
\affiliation{
\institution{National University of Singapore}
    \city{Singapore}
  \country{Singapore}
  }

\author{Xin Yi}
\orcid{0000-0001-8041-7962}
\email{yixin@tsinghua.edu.cn}
\affiliation{
    \department{Institute for Network Sciences and Cyberspace}
    \institution{Tsinghua University}
    \city{Beijing}
    \country{China}
}

\author{Jorge Goncalves}
\email{jorge.goncalves@unimelb.edu.au}
\orcid{0000-0002-0117-0322}
\affiliation{%
 \department{School of Computing and Information Systems}
 \institution{University of Melbourne}
 \city{Melbourne}
  \country{Australia}
  }

\renewcommand{\shortauthors}{Ying Ma et al.}


\begin{abstract}
AI-powered lifelogging wearables are emerging as a new class of consumer devices that transform everyday experience into searchable, AI-curated memory archives. We study early public sensemaking around these systems at the moment of their market entry, using the Looki L1 as an empirical lens. Analysing large-scale Chinese-language and English-language social media discourse (N = 5,053 comments), we combine topic clustering with inductive thematic analysis to examine how users interpret the social, moral, and political implications of AI-mediated memory. Across contexts, users reference dystopian surveillance imaginaries, express privacy resignation and bystander concerns, and debate assistive value alongside consumer logics. English-language comments more often framed these devices through interpersonal power, evidentiary use, and hacking anxieties, while Chinese-language comments more often foregrounded labour exploitation, governance surveillance, and technological inevitability. 


\end{abstract}




\maketitle

\section{Introduction}

Wearable cameras and smart glasses have long been explored as a promising direction for ubiquitous and embodied computing. From early lifelogging devices to contemporary smart glasses, prior systems have largely focused on episodic first-person capture, enabling users to manually record photos or videos for later review and sharing~\cite{denning2014situ,hu2023investigating}. Research in HCI has extensively examined these devices in terms of social acceptability, bystander privacy, recording norms, and everyday use practices \cite{bhardwaj2024focus,zhou2024social}.

Recently, a new class of AI-powered wearable cameras has begun to emerge, shifting wearable capture from manual recording toward ambient sensing and automated interpretation~\cite{hodges2006sensecam}. Rather than functioning solely as personal recording tools, these devices increasingly integrate onboard sensing, cloud-based inference, and generative AI, transforming visual data into structured, searchable, and narrativised representations of everyday life~\cite{gurrin2013exploring,ma2025raising}. The Looki L1 \footnote{https://www.looki.ai/} represents a consequential example of this transition. It is marketed as an AI lifelogging camera and proactive AI companion. Unlike traditional smart glasses or wearable cameras that rely on user-initiated recording, Looki operates through an event-driven “Story Mode” that selectively captures short clips when it detects meaningful changes in a user’s environment or activity. These fragments are curated by generative AI into daily summaries and transformed into a searchable memory archive through a Retrieval-Augmented Generation (RAG) pipeline, enabling users to query their recent past in natural language~\cite{lewis2020retrieval,zhang2025understanding}.

What's more, Looki L1 entered the consumer market in August 2025, triggering a wave of influencer reviews and user-generated evaluation posts across platforms such as TikTok, Instagram and Xiaohongshu, etc. 
Prior research suggests that social media discourse provides a valuable empirical lens for examining how publics collectively interpret and negotiate the meaning of emerging technologies as they are introduced into everyday life \cite{ma2023hello,lambert2025does,choi2025creator,zhang2025dark,gamage2022deepfakes}. 
As such, these posts and their accompanying comment threads provide a rare opportunity to capture users’ first encounters with an AI-powered wearable device in real-world contexts. Analysing this discourse allows us to observe naturalistic reactions, concerns, and expectations as they emerge in situ, beyond controlled or researcher-mediated settings.
Therefore, we pose the following research questions:

\textbf{RQ1}: How do users on social media perceive AI-powered  lifelogging wearables at the moment of their emergence?

\textbf{RQ2}: What shared and context-specific discourses emerge across Chinese-language (CL) and English-language (EL) social media comments?

This work contributes an empirical study of early public framings of AI lifelogging wearables, a comparative analysis of CL and EL social media discourse that distinguishes shared anxieties from context-specific imaginaries, and design and governance implications for consent, accountability, and culturally situated data practices as AI-mediated lifelogging systems enter everyday life.

\section{Background: Smart Glasses and Wearable Cameras in Everyday Life}

The key distinction between smart glasses and wearable cameras lies in sensing perspective and interaction agency. Smart glasses are gaze-aligned, capturing a high-frequency, subjective view of user attention and supporting real-time interaction as “eyewear computers” \cite{bulling2016eyewear}. Wearable cameras, in contrast, are typically torso-aligned, providing a more stable environmental perspective that is well suited for long-form activity recognition and lifelogging \cite{gurrin2014lifelogging}. Beyond hardware placement, these devices also differ in agency: smart glasses operate as reactive assistants triggered by user commands, whereas wearable cameras enable proactive, always-on sensing systems that automatically summarise daily experience and support continuous memory augmentation \cite{fang2025mirai, lee2023memoro}.

HCI research has long examined the everyday use of smart glasses and wearable cameras, focusing on social acceptability, bystander reactions, and privacy tensions in public and semi-public spaces \cite{zhang2025situguard}. Early field studies of Google Glass showed that bystanders respond with curiosity, discomfort, and resistance, often driven by uncertainty about whether recording is taking place and what data are being captured \cite{denning2014situ}. Subsequent work on lifelogging demonstrates that wearable cameras routinely capture sensitive and intimate moments, with seemingly mundane activities revealing highly personal information about relationships, routines, and locations \cite{hoyle2015sensitive}. Studies of lifeloggers’ practices further show how users develop ad hoc strategies to manage social awkwardness and mitigate privacy concerns when wearing always-on cameras in shared environments \cite{hoyle2014privacy}.

As consumer camera glasses have re-emerged in recent years (e.g., Ray-Ban Meta \footnote{https://www.meta.com/au/ai-glasses/}, Snap Spectacles \footnote{https://www.spectacles.com/}), previous research has revisited these issues in the context of contemporary social media and platform ecosystems. Recent CHI work examining camera glasses from the wearer’s perspective shows that users experience persistent social friction when navigating everyday interactions, including discomfort when recording around strangers, uncertainty about appropriate use contexts, and concerns about how others perceive the device \cite{bhardwaj2024focus}. These studies suggest that public visibility and ambiguity about device capabilities continue to shape the social acceptability of smart glasses, even as their form factors become more discreet and normalised.

This work positions smart glasses and wearable cameras as recurring sociotechnical artefacts that repeatedly surface unresolved tensions between personal utility and social accountability. Rather than being isolated technological novelties, such devices participate in long-standing cycles of hype, backlash, and redesign, where everyday practices become sites of negotiation over legitimacy, trust, and appropriateness. This historical continuity provides important context for understanding contemporary reactions to AI-powered lifelogging wearables (AI-PLWs) such as Looki, which build upon earlier camera-based wearables while introducing new forms of automated perception and inference \cite{zhang2026pervasive}.

\section{Method}


\subsection{Data Collection}

We collected public posts and comments related to Looki AI from social media platforms containing both CL and EL content, including TikTok, Xiaohongshu, Instagram and YouTube. Data was gathered using a combination of official platform APIs (where available) and custom web scrapers for platforms without public APIs.
We excluded platforms that identified less than 100 posts.
Namely, our keyword search identified less than 100 posts/replies on X and Reddit respectively. In addition, the posts from official Looki.ai corporate accounts were removed from the dataset across all scraped social media platforms to minimise the risk of comment manipulation.
We conducted a keyword-based search to select relevant posts discussing the Looki AI product using the query “looki” and required a match with at least one of the following keywords: “looki”, “looki.ai”, “lookiL1”, “L1”, “camera”, “vlog”, “AI”, or “wearable.” Following manual screening to filter out irrelevant posts, we collected all posts and reply threads across each platform.
We collected posts, their associated comments, timestamps, and engagement metadata from August 2025 to December 2025. We retained only the comment text and platform-level metadata required for analysis and did not collect any identifying information. After removing unrelated posts, the final dataset comprises 2,711 EL comments and 2,342 CL comments.

\subsection{Data Processing}

We adopt a machine-learning–assisted approach to identify quantitative themes, followed by human-driven inductive analysis on a sample of posts from each theme. 
We use topic-clustering methods (BERTopic and LDA) to surface both prevalent and less prominent themes~\cite{blei2003latent, bertopic_grootendorst22}. 
Topic clusters were generated separately for CL and EL comments to avoid merging discourse shaped by different languages, platforms, and audience contexts.
Researchers then conducted an inductive thematic analysis within each topic cluster \cite{fereday2006demonstrating}. This involved iteratively reading and open-coding a sample of posts from each cluster to develop a contextualised and interpretive understanding of the emergent themes. The first author iteratively read and open-coded a sample of comments from each cluster, which two co-authors reviewed; ambiguous cases were resolved through discussion. 
Through this process, we refined topic labels, identified recurring discursive patterns, and synthesised higher-level thematic categories grounded in the data.

\section{Results}

We identify both shared and language-situated discourses around AI-PLWs, as shown in Figure 1. Some themes appear across both CL and EL comments (Sec. \ref{shared}), while others appear more prominently within EL comments (Sec. \ref{Western}) or CL discussions (Sec. \ref{Chinese}).

\begin{figure}[htbp]
  \includegraphics[width= 0.46\textwidth]{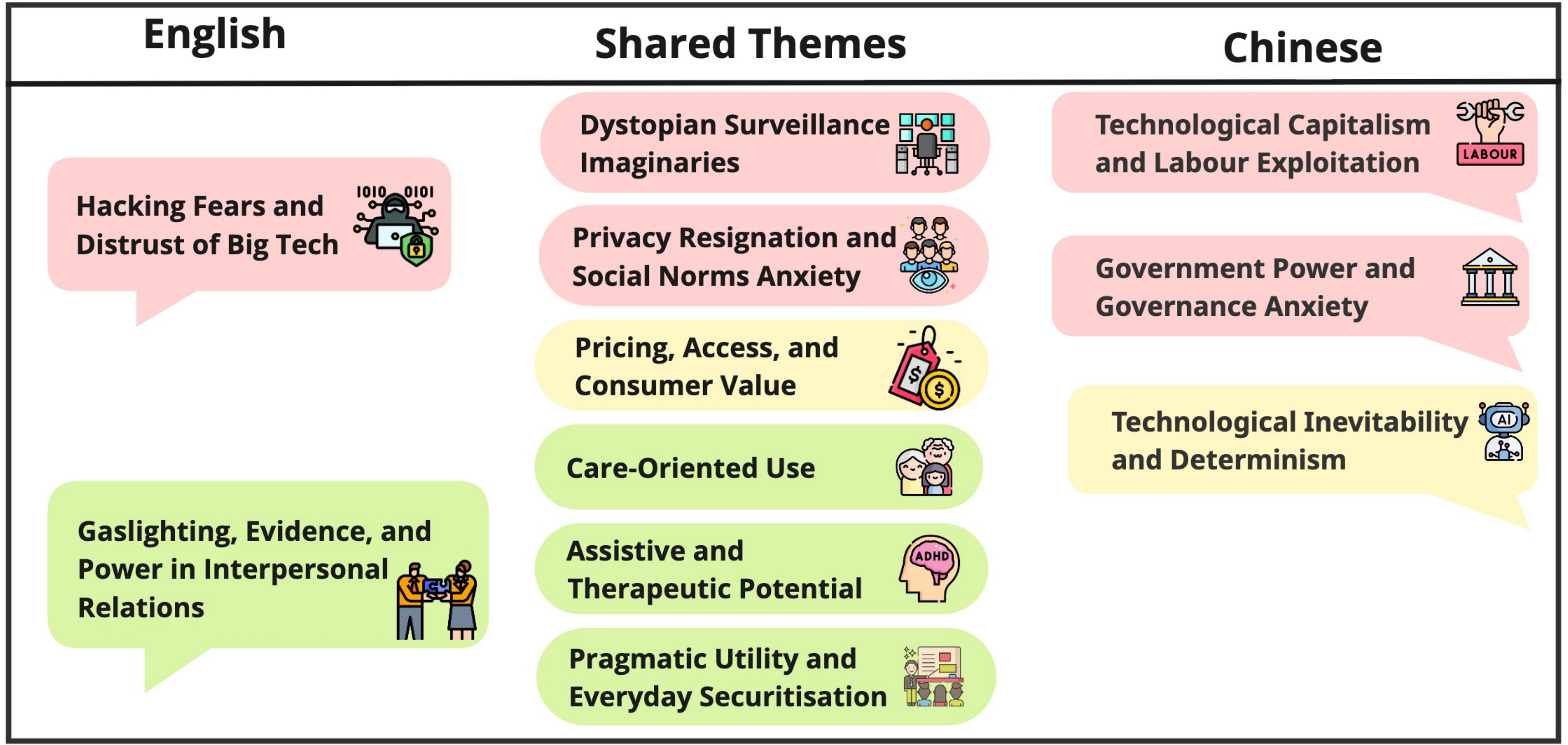}
  \caption{Shared and language-situated public discourses on AI-powered lifelogging wearables}
\Description{
A three-column diagram comparing public discourse themes about AI-powered lifelogging wearables in English- and Chinese-language discussions. The columns are labelled ``English'' on the left, ``Shared Themes'' in the centre, and ``Chinese'' on the right. The centre column contains six themes that appeared across both language contexts. From top to bottom, these are: ``Dystopian Surveillance Imaginaries''; ``Privacy Resignation and Social Norms Anxiety''; ``Pricing, Access, and Consumer Value''; ``Care-Oriented Use''; ``Assistive and Therapeutic Potential''; and ``Pragmatic Utility and Everyday Securitisation.'' 

The English column contains two themes that were more specific or salient in the English-language discourse: ``Hacking Fears and Distrust of Big Tech'' near the upper left, and ``Gaslighting, Evidence, and Power in Interpersonal Relations'' near the lower left. The Chinese column contains three themes that were more specific or salient in the Chinese-language discourse: ``Technological Capitalism and Labour Exploitation'' near the upper right, ``Government Power and Governance Anxiety'' below it, and ``Technological Inevitability and Determinism'' in the middle-to-lower right.}
  \label{theme}
\end{figure}



\subsection{Shared Themes}
\label{shared}


\textbf{Dystopian Surveillance Imaginaries}:
A large subset of users in both datasets interpret AI-PLWs through dystopian cultural references, most notably Black Mirror, \footnote{https://en.wikipedia.org/wiki/Black\_Mirror}, 1984 \footnote{https://en.wikipedia.org/wiki/Nineteen\_Eighty-Four}, and “Big Brother.” Rather than engaging with specific technical affordances, commenters situate these systems within broader narratives of technological overreach and moral decline.
Representative comments include: “Straight out of Black Mirror,” and “1984 wasn’t supposed to be a product roadmap.” 
These comments frame wearable lifelogging as a symbol of voluntary surveillance, where privacy loss is no longer imposed but embraced. The technology becomes a metaphor for a future in which continuous monitoring is normalised and resistance is futile.

\textbf{ Assistive and Therapeutic Potential:} 
A substantial subset of users highlight the assistive potential of AI-PLWs, particularly for people with dementia, memory loss, cognitive impairment, and disability. These users frame such systems as cognitive prostheses: external memory infrastructures that can support daily functioning, independence, and safety.
Representative comments include: “Great for people with dementia.” “This could help Alzheimer’s patients.” “Perfect for ADHD.” 


\textbf{Privacy Resignation and Social Norms Anxiety}: Some users express a sense of privacy resignation, framing surveillance as an unavoidable condition of contemporary digital life. Rather than resisting monitoring, commenters suggest that privacy has already been irretrievably lost, stating that “we gave up privacy years ago” and “they already have everything.” This discourse reflects the normalisation of surveillance, where consent becomes symbolic rather than substantive.
At the same time, users articulate strong anxieties around bystander consent, covert recording, legal responsibility, and social conflict. AI-PLWs are framed as socially illegitimate, with users asking “Wouldn't it be illegal to record in a public area without their consent?” The central concern is not technical feasibility, but whether public recording violates unspoken norms of visibility, consent, and moral legitimacy in shared spaces.







\textbf{Care-Oriented Use}:
Some users frame AI-PLWs through a care-oriented and utilitarian lens, emphasising their potential value for elder care, child safety, and everyday support.
Users imagine these systems being deployed in institutional and domestic settings, such as nursing homes and kindergartens, suggesting that “this could be used in nursing homes,” “it would be good for elderly people,” and “this could be used in kindergartens.”


\textbf{Pragmatic Utility and Everyday Securitisation}: Some users frame these devices through a pragmatic lens, emphasising their everyday usefulness for learning, documentation, and memory offloading. Users describe using the device “at school to record discussions” and “no longer needing to write a diary,” positioning wearable AI as a productivity-enhancing tool embedded in daily routines.
A related interpretive frame positions these devices as mobile security infrastructures analogous to home CCTV systems, dashcams, or body cameras. Users imagine wearable AI as a deterrent against crime and a tool for personal protection, describing it as “basically a body cam,” and “good for evidence if something happens.” This discourse reflects the normalisation of continuous recording as a form of everyday securitisation.


\textbf{Pricing, Access, and Consumer Value}:
Finally, many users assess AI-PLWs through everyday consumer logics, questioning affordability, battery life, data storage, and ongoing fees. Representative comments include: “Too expensive for what it does,” “Only rich people will buy this,” “How can the battery last so long?” “Where is all the data stored?” and “Is there a monthly subscription?”

\subsection{Themes in English-Language Comments}
\label{Western}
\textbf{Gaslighting, Evidence, and Power in Interpersonal Relations}
Some users framed AI-powered wearable devices as “evidence machines” that can mediate disputes, resolve conflicts, and counteract manipulation or gaslighting. In these comments, the device is imagined as an externalised witness that redistributes epistemic authority between people by providing an objective record of past interactions.
Representative comments include: “Bad news for gaslighters.” “Anti-gaslighting tech.” “The amount of arguments I could win.” “Narcissists are in trouble.”
Users frequently describe scenarios involving partners, friends, or colleagues who deny past statements or events, positioning wearable AI as a tool for reclaiming credibility and protecting oneself from psychological manipulation. This discourse reflects growing concerns around interpersonal accountability in digitally mediated life and positions lifelogging systems not only as memory aids but also as mechanisms for enforcing social truth and power symmetry.

\textbf{Hacking Fears and Distrust of Big Tech}: 
Another prominent interpretive frame centres on fears of hacking, data misuse, and loss of control. Users imagine scenarios in which body-worn cameras are stolen, remotely accessed, or secretly exploited by third parties. The presence of an always-on sensing device generates a sense of permanent exposure, with commenters warning, “What if someone hacks your camera?” and “You lose it and now someone has your whole life.” Others describe the device as “basically wearing a spy on your body.”
A related discourse reflects deep distrust toward large technology companies and cloud infrastructures, with several users explicitly questioning the safety of AWS and expressing scepticism toward centralised data storage. In this framing, wearable AI is not understood as empowering, but as a technological vulnerability that renders the body open to digital intrusion.

\subsection{Themes in Chinese-Language Comments}
\label{Chinese}
\textbf{Technological Capitalism and Labour Exploitation}:
Some users frame AI-PLWs through a political-economic lens, interpreting them as tools for intensifying labour control and managerial surveillance. Users describe the technology as “a gift to capitalists,” while workers remain trapped in exhausting routines. Others warn of “a return to the age of slavery” and describe workers as being “squeezed dry.”
In this discourse, wearable AI is not imagined as a personal assistant, but as an extension of managerial power and extractive platform capitalism.

\textbf{Government Power and Governance Anxiety}: Some users extend surveillance anxieties to the level of the government, imagining AI-PLWs as precursors to mandatory tracking systems, digital identity, or algorithmic governance. 
Commenters imagine “a future of permanent citizen records”, “big-data society”, and “an upgraded social credit system”. This discourse reflects a political reading of AI surveillance as embedded in state power and population management.

\textbf{Technological Inevitability and Determinism}:
Many users express a deterministic view of technological progress, describing AI surveillance as “unstoppable”, “inevitable”, and “a broader trend of the times”.

\section{Discussion}

Our findings suggest that public responses to AI-PLWs are not simply repetitions of earlier concerns about wearable cameras or smart glasses. While familiar issues of privacy, surveillance, and social acceptability remain central, AI changes the nature of these concerns by transforming captured moments into searchable, interpretable, and potentially evidentiary records. Commenters were not only concerned about being recorded; they were also concerned about how AI might summarise events, infer intentions, reconstruct memories, and make everyday interactions available for later review. However, this study focuses on a single product, and its findings may not generalise to other AI-PLWs.

\subsection{The Moral Tension of AI-powered Memory Wearables}
Across platforms, users consistently interpret AI-powered memory wearables as more than recording devices. Rather than focusing on technical features, public discourse situates these systems within broader narratives of technological overreach, social control, and moral risk. On the one hand, users envision wearable memory devices as compassionate companions for vulnerable populations. On the other, they fear that the same infrastructures enable unprecedented forms of bodily monitoring and behavioural capture. These competing framings position AI memory wearables within a moral tension between care and surveillance.
AI-powered memory wearables also differ from earlier generations of continuous-sensing devices such as lifelogging cameras. They are more generalised and better suited to a wider range of tasks~\cite{korayem2016enhancing}, and are more subtle in form, making them harder to detect in everyday interactions~\cite{phelan2016s}.
This tension suggests that the social legitimacy of AI memory wearables will depend not only on their functional benefits, but on whether they can convincingly demonstrate that care-oriented uses are not subordinated to surveillance and data extraction. Public acceptance is therefore shaped as much by the surrounding data infrastructure as by the device itself.


\subsection{Cultural Visions of AI-powered Memory Wearables}

The divergence between EL and CL comments suggests different discourse framings of AI-PLWs as sociotechnical objects. In EL comments, these devices are primarily framed through interpersonal relations and individual agency. The recurring image of AI memory as an “evidence machine” suggests that memory is understood as a resource for negotiating trust, accountability, and power within everyday relationships. 
In contrast, Chinese discourse situates AI memory wearables within a broader political-economic imaginary. Rather than focusing on individual disputes, users interpret these systems as extensions of platform capitalism, labour management, and large-scale data infrastructures. Memory is framed less as a personal tool and more as part of organisational and societal systems of optimisation. The emphasis on technological inevitability reflects a broader cultural narrative in which technological development is perceived as an unstoppable historical trend. This framing reflects \textit{holistic thinking}~\cite{nisbett2001culture} combined with higher \textit{power distance}~\cite{hofstede2011dimensionalizing}, situating the device within broader platform economies and labour hierarchies. Here, the technology is not merely a personal aid but a tool through which supervisory power is enacted.

These contrasting framings suggest that AI memory wearables are likely to be socially positioned either as personal cognitive tools or as societal data infrastructures, depending on cultural context. This difference has important consequences for how responsibility, control, and acceptable use are publicly negotiated.

Accordingly, designers and policymakers should not assume a universal model of user expectations for AI memory wearables~\cite{wilkinson2018moving}. Instead, governance, consent mechanisms, and accountability structures must be culturally situated.

\bibliographystyle{ACM-Reference-Format}
\balance
\bibliography{sample-base}

@String{Computing = "Computing" }

@String{Springer = "Springer-Verlag" }

@inproceedings{choi2025creator,
  title={Creator Hearts: Investigating the Impact Positive Signals from YouTube Creators in Shaping Comment Section Behavior},
  author={Choi, Frederick and Lambert, Charlotte and Koshy, Vinay and Pratipati, Sowmya and Do, Tue and Chandrasekharan, Eshwar},
  booktitle={Proceedings of the 2025 CHI Conference on Human Factors in Computing Systems},
  pages={1--18},
  year={2025}
}

@inproceedings{lambert2025does,
  title={Does Positive Reinforcement Work?: A Quasi-Experimental Study of the Effects of Positive Feedback on Reddit},
  author={Lambert, Charlotte and Saha, Koustuv and Chandrasekharan, Eshwar},
  booktitle={Proceedings of the 2025 CHI Conference on Human Factors in Computing Systems},
  pages={1--16},
  year={2025}
}

@inproceedings{gamage2022deepfakes,
  title={Are deepfakes concerning? analyzing conversations of deepfakes on reddit and exploring societal implications},
  author={Gamage, Dilrukshi and Ghasiya, Piyush and Bonagiri, Vamshi and Whiting, Mark E and Sasahara, Kazutoshi},
  booktitle={Proceedings of the 2022 CHI conference on human factors in computing systems},
  pages={1--19},
  year={2022}
}

@inproceedings{zhang2025dark,
  title={The dark side of ai companionship: A taxonomy of harmful algorithmic behaviors in human-ai relationships},
  author={Zhang, Renwen and Li, Han and Meng, Han and Zhan, Jinyuan and Gan, Hongyuan and Lee, Yi-Chieh},
  booktitle={Proceedings of the 2025 CHI Conference on Human Factors in Computing Systems},
  pages={1--17},
  year={2025}
}

@inproceedings{ma2023hello,
  title={“Hello, fellow villager!”: Perceptions and impact of displaying users’ locations on weibo},
  author={Ma, Ying and Zhou, Qiushi and Tag, Benjamin and Sarsenbayeva, Zhanna and Knibbe, Jarrod and Goncalves, Jorge},
  booktitle={IFIP Conference on Human-Computer Interaction},
  pages={511--532},
  year={2023},
  organization={Springer}
}

@inproceedings{denning2014situ,
  title={In situ with bystanders of augmented reality glasses: Perspectives on recording and privacy-mediating technologies},
  author={Denning, Tamara and Dehlawi, Zakariya and Kohno, Tadayoshi},
  booktitle={Proceedings of the SIGCHI conference on human factors in computing systems},
  pages={2377--2386},
  year={2014}
}

@inproceedings{hoyle2015sensitive,
  title={Sensitive lifelogs: A privacy analysis of photos from wearable cameras},
  author={Hoyle, Roberto and Templeman, Robert and Anthony, Denise and Crandall, David and Kapadia, Apu},
  booktitle={Proceedings of the 33rd Annual ACM conference on human factors in computing systems},
  pages={1645--1648},
  year={2015}
}

@inproceedings{hoyle2014privacy,
  title={Privacy behaviors of lifeloggers using wearable cameras},
  author={Hoyle, Roberto and Templeman, Robert and Armes, Steven and Anthony, Denise and Crandall, David and Kapadia, Apu},
  booktitle={Proceedings of the 2014 ACM international joint conference on pervasive and ubiquitous computing},
  pages={571--582},
  year={2014}
}

@inproceedings{bhardwaj2024focus,
  title={In Focus, Out of Privacy: The Wearer's Perspective on the Privacy Dilemma of Camera Glasses},
  author={Bhardwaj, Divyanshu and Ponticello, Alexander and Tomar, Shreya and Dabrowski, Adrian and Krombholz, Katharina},
  booktitle={Proceedings of the 2024 CHI Conference on Human Factors in Computing Systems},
  pages={1--18},
  year={2024}
}

@article{blei2003latent,
  title={Latent dirichlet allocation},
  author={Blei, David M and Ng, Andrew Y and Jordan, Michael I},
  journal={Journal of machine Learning research},
  volume={3},
  number={Jan},
  pages={993--1022},
  year={2003}
}

@article{bertopic_grootendorst22,
  title={BERTopic: Neural topic modeling with a class-based TF-IDF procedure},
  author={Grootendorst, Maarten},
  journal={arXiv preprint arXiv:2203.05794},
  year={2022}
}

@article{bulling2016eyewear,
author = {Bulling, Andreas and Kunze, Kai},
title = {Eyewear computers for human-computer interaction},
year = {2016},
issue_date = {May + June 2016},
publisher = {Association for Computing Machinery},
address = {New York, NY, USA},
volume = {23},
number = {3},
issn = {1072-5520},
url = {https://doi.org/10.1145/2912886},
doi = {10.1145/2912886},
journal = {Interactions},
month = apr,
pages = {70–73},
numpages = {4}
}

@article{gurrin2014lifelogging,
    author = {Gurrin, Cathal and Smeaton, Alan F. and Doherty, Aiden R.},
    title = {LifeLogging: Personal Big Data},
    journal = {Foundations and Trends in Information Retrieval},
    volume = {8},
    number = {1},
    pages = {1-125},
    year = {2014},
    month = {06},
    issn = {1554-0669},
    doi = {10.1561/1500000033},
    url = {https://doi.org/10.1561/1500000033},
    eprint = {https://www.emerald.com/ftinr/article-pdf/8/1/1/11048006/1500000033en.pdf},
}

@inproceedings{fang2025mirai,
author = {Fang, Cathy Mengying and Samaradivakara, Yasith and Maes, Pattie and Nanayakkara, Suranga},
title = {Mirai: A Wearable Proactive AI "Inner-Voice" for Contextual Nudging},
year = {2025},
isbn = {9798400713958},
publisher = {Association for Computing Machinery},
address = {New York, NY, USA},
url = {https://doi.org/10.1145/3706599.3719881},
doi = {10.1145/3706599.3719881},
booktitle = {Proceedings of the Extended Abstracts of the CHI Conference on Human Factors in Computing Systems},
articleno = {399},
numpages = {9},
location = {
},
series = {CHI EA '25}
}

@inproceedings{lee2023memoro,
author = {Zulfikar, Wazeer Deen and Chan, Samantha and Maes, Pattie},
title = {Memoro: Using Large Language Models to Realize a Concise Interface for Real-Time Memory Augmentation},
year = {2024},
isbn = {9798400703300},
publisher = {Association for Computing Machinery},
address = {New York, NY, USA},
url = {https://doi.org/10.1145/3613904.3642450},
doi = {10.1145/3613904.3642450},
booktitle = {Proceedings of the 2024 CHI Conference on Human Factors in Computing Systems},
articleno = {450},
numpages = {18},
location = {Honolulu, HI, USA},
series = {CHI '24}
}

@inproceedings{hodges2006sensecam,
  title={SenseCam: A retrospective memory aid},
  author={Hodges, Steve and Williams, Lyndsay and Berry, Emma and Izadi, Shahram and Srinivasan, James and Butler, Alex and Smyth, Gavin and Kapur, Narinder and Wood, Ken},
  booktitle={International conference on ubiquitous computing},
  pages={177--193},
  year={2006},
  organization={Springer}
}

@article{hu2023investigating,
  title={Investigating the Design Considerations for Integrating Text-to-Image Generative AI within Augmented Reality Environments},
  author={Hu, Yongquan and Zhang, Dawen and Yuan, Mingyue and Xian, Kaiqi and Elvitigala, Don Samitha and Kim, June and Mohammadi, Gelareh and Xing, Zhenchang and Xu, Xiwei and Quigley, Aaron},
  journal={arXiv preprint arXiv:2303.16593},
  year={2023}
}

@inproceedings{zhou2024social,
  title={Social xr: Designing an extended reality application for interaction in long-distance relationships},
  author={Zhou, Dezijian and Deng, Yuanyuan and Globa, Anastasia and Davies, Richard and Hu, Yongquan'Owen and Wu, Ruo-Xuan and Fu, Liya and Quigley, Aaron},
  booktitle={Proceedings of the 36th Australasian Conference on Human-Computer Interaction},
  pages={844--852},
  year={2024}
}

@article{lewis2020retrieval,
  title={Retrieval-augmented generation for knowledge-intensive nlp tasks},
  author={Lewis, Patrick and Perez, Ethan and Piktus, Aleksandra and Petroni, Fabio and Karpukhin, Vladimir and Goyal, Naman and K{\"u}ttler, Heinrich and Lewis, Mike and Yih, Wen-tau and Rockt{\"a}schel, Tim and others},
  journal={Advances in neural information processing systems},
  volume={33},
  pages={9459--9474},
  year={2020}
}

@inproceedings{gurrin2013exploring,
  title={Exploring the technical challenges of large-scale lifelogging},
  author={Gurrin, Cathal and Smeaton, Alan F and Qiu, Zhengwei and Doherty, Aiden},
  booktitle={Proceedings of the 4th international SenseCam \& pervasive imaging conference},
  pages={68--75},
  year={2013}
}

@article{fereday2006demonstrating,
  title={Demonstrating rigor using thematic analysis: A hybrid approach of inductive and deductive coding and theme development},
  author={Fereday, Jennifer and Muir-Cochrane, Eimear},
  journal={International journal of qualitative methods},
  volume={5},
  number={1},
  pages={80--92},
  year={2006},
  publisher={SAGE Publications Sage CA: Los Angeles, CA}
}

@inproceedings{wilkinson2018moving,
  title={Moving beyond a" one-size fits all" exploring individual differences in privacy},
  author={Wilkinson, Daricia and Namara, Moses and Badillo-Urquiola, Karla and Wisniewski, Pamela J and Knijnenburg, Bart P and Page, Xinru and Toch, Eran and Romano-Bergstrom, Jen},
  booktitle={Extended Abstracts of the 2018 CHI Conference on Human Factors in Computing Systems},
  pages={1--8},
  year={2018}
}

@inproceedings{korayem2016enhancing,
  title={Enhancing lifelogging privacy by detecting screens},
  author={Korayem, Mohammed and Templeman, Robert and Chen, Dennis and Crandall, David and Kapadia, Apu},
  booktitle={Proceedings of the 2016 CHI Conference on Human Factors in Computing Systems},
  pages={4309--4314},
  year={2016}
}

@inproceedings{phelan2016s,
  title={It's creepy, but it doesn't bother me},
  author={Phelan, Chanda and Lampe, Cliff and Resnick, Paul},
  booktitle={Proceedings of the 2016 CHI conference on human factors in computing systems},
  pages={5240--5251},
  year={2016}
}

@article{nisbett2001culture,
  title={Culture and systems of thought: holistic versus analytic cognition.},
  author={Nisbett, Richard E and Peng, Kaiping and Choi, Incheol and Norenzayan, Ara},
  journal={Psychological review},
  volume={108},
  number={2},
  pages={291},
  year={2001},
  publisher={American Psychological Association}
}

@article{hofstede2011dimensionalizing,
  title={Dimensionalizing cultures: The Hofstede model in context},
  author={Hofstede, Geert},
  journal={Online readings in psychology and culture},
  volume={2},
  number={1},
  pages={8},
  year={2011},
  publisher={International Association for Cross-Cultural Psychology}
}

@inproceedings{ma2025raising,
  title={Raising Awareness of Location Information Vulnerabilities in Social Media Photos using LLMs},
  author={Ma, Ying and Zhang, Shiquan and Yang, Dongju and Sarsenbayeva, Zhanna and Knibbe, Jarrod and Goncalves, Jorge},
  booktitle={Proceedings of the 2025 CHI Conference on Human Factors in Computing Systems},
  pages={1--14},
  year={2025}
}

@article{zhang2026pervasive,
  title={The Pervasive Blind Spot: Benchmarking VLM Inference Risks on Everyday Personal Videos},
  author={Zhang, Shuning and Li, Zhaoxin and Wen, Changxi and Ma, Ying and Li, Simin and Zhang, Gengrui and Zhang, Ziyi and Meng, Yibo and Zhao, Hantao and Yi, Xin and others},
  journal={Proceedings of the ACM on Interactive, Mobile, Wearable and Ubiquitous Technologies},
  volume={10},
  number={2},
  pages={1--38},
  year={2026},
  publisher={ACM New York, NY, USA}
}

@inproceedings{zhang2025situguard,
  title={SituGuard: LLM-based Fine-grained Smart Glass Privacy Control in Home Environments},
  author={Zhang, Shuning and Ma, Ying and Zang, Qucheng and Hu, Yongquan'Owen'},
  booktitle={Companion of the 2025 ACM International Joint Conference on Pervasive and Ubiquitous Computing},
  pages={609--612},
  year={2025}
}

@inproceedings{zhang2025understanding,
  title={Understanding Users' Privacy Perceptions Towards LLM's RAG-based Memory},
  author={Zhang, Shuning and Ma, Rongjun and Ma, Ying and Li, Shixuan and Xu, Yiqun and Yi, Xin and Li, Hewu},
  booktitle={Proceedings of the 2025 Workshop on Human-Centered AI Privacy and Security},
  pages={10--19},
  year={2025}
}

\end{document}